\documentclass[sigconf]{acmart}

\copyrightyear{2026}
\acmYear{2026}
\setcopyright{cc}
\setcctype{by}
\acmConference[CCS '26] {Proceedings of the 2026 ACM SIGSAC Conference on Computer and Communications Security}{November 15--19, 2026}{The Hague, Netherlands.}
\acmBooktitle{Proceedings of the 2026 ACM SIGSAC Conference on Computer and Communications Security (CCS '26), November 15--19, 2026, The Hague, Netherlands}
\acmISBN{979-8-4007-2871-6/2026/11}
\acmDOI{10.1145/3830454.3846541}

\usepackage{enumitem}
\usepackage{colortbl}
\usepackage{xspace}
\usepackage{balance}

\definecolor{lightred}{RGB}{255,200,200}

\newcommand{\BfPara}[1]{{\noindent\bf#1.}\xspace}
\newcommand{\eg}{e.g.,\@\xspace}
\newcommand{\ie}{i.e.,\@\xspace}
\newcommand{\etal}{et~al.\@\xspace}

\newcommand{\takeaway}[1]{%
  \par\smallskip
  \begingroup
  \setlength{\fboxsep}{1mm}%
  \setlength{\fboxrule}{1pt}%
  \noindent
  \fcolorbox{black}{gray!7!white}{%
    \parbox{\dimexpr\columnwidth-2\fboxsep-2\fboxrule\relax}{%
      \textbf{Takeaway:} #1%
    }%
  }%
  \endgroup
  \smallskip
}

\begin{document}

\title{The Tragedy of Convenience: Cascading User-Data Leakage from SMS-Delivered URLs}

\author{Muhammad Danish}
\orcid{0009-0007-9634-3540}
\affiliation{
  \institution{University of New Mexico}
  \city{Albuquerque}
  \state{New Mexico}
  \country{USA}
  }
\email{mdanish@unm.edu}

\author{Enrique Sobrados}
\orcid{0009-0009-6864-3911}
\affiliation{
  \institution{University of New Mexico}
  \city{Albuquerque}
  \state{New Mexico}
  \country{USA}
  }
\email{esobrados720@unm.edu}

\author{Priya Kaushik}
\orcid{0000-0002-0427-3431}
\affiliation{
  \institution{University of Arizona}
  \city{Tucson}
  \state{Arizona}
  \country{USA}
  }
\email{priyakaushik@arizona.edu}

\author{Bhupendra Acharya}
\orcid{0009-0005-3663-152X}
\affiliation{
  \institution{University of Louisiana at Lafayette}
  \city{Lafayette}
  \state{Louisiana}
  \country{USA}
  }
\email{bhupendra.acharya@louisiana.edu}

\author{Muhammad Saad}
\orcid{0000-0002-8762-4566}
\affiliation{
  \institution{Circle Internet Group}
  \city{Boston}
  \state{Massachusetts}
  \country{USA}
  }
\email{saad@circle.com}

\author{Abdullah Mueen}
\orcid{0000-0002-4839-1624}
\affiliation{
  \institution{University of New Mexico}
  \city{Albuquerque}
  \state{New Mexico}
  \country{USA}
  }
\email{mueen@unm.edu}

\author{Sazzadur Rahaman}
\orcid{0000-0002-1258-6470}
\affiliation{
  \institution{University of Arizona}
  \city{Tucson}
  \state{Arizona}
  \country{USA}
  }
\email{sazz@arizona.edu}

\author{Afsah Anwar}
\orcid{0000-0003-1449-3590}
\affiliation{
  \institution{University of New Mexico}
  \city{Albuquerque}
  \state{New Mexico}
  \country{USA}
  }
\email{afsah@unm.edu}

\renewcommand{\shortauthors}{Danish et al.}

\begin{abstract}
Digital Services are increasingly sending private URLs over Short Message Service (SMS) to allow users to resume sessions with a single click. While convenient, this design shifts trust from explicit authentication to a potentially vulnerable communication channel. Basically, the vulnerability lies in the assumption that the private link can only be accessed by the intended user.

In this paper, we demonstrate that this assumption can be easily violated. In particular, we show how seemingly isolated link exposure can cascade into a wider data leak. Using public SMS gateways as an ethical lens, we analyze more than 322K unique private URLs extracted from over 33 million messages across 30K+ phone numbers. Across 701 URLs, we find that at least 177 web services effectively treat private URLs as bearer credentials, enabling unauthorized access to sensitive user information (e.g., financial details, national IDs) once the link is exposed.

Alarmingly, we show that 125 services are potentially enumerable, i.e., a single URL can lead to a cascading effect, resulting in the data leakage of their entire user base. Moreover, we observe that 5 services that implement authentication partially reveal account information before authentication is completed and rely on lightweight parameters (e.g., date of birth, ZIP code).
Even worse, in 4 out of these 5 services, the authentication is vulnerable to brute-force attacks. Further, we uncover that 20 services grant privileged access: 14 allow modification of Personally Identifiable Information (PII), 5 grant account access, and 1 allows both. We also find 84 services that expose additional PII beyond the landing page. Our disclosures led to acknowledgments from 18 services, 7 of which have already been fixed, positively impacting at least 120 million users. 
\end{abstract}



\maketitle

\section{Introduction}
\label{sec:short_introduction}

Increasing awareness of web security threats has driven the adoption of stronger authentication mechanisms~\cite{gavazzi2023study, lassak2024aren}. Stricter password policies and Multi-Factor Authentication (MFA) have improved resistance to attacks such as credential stuffing and phishing. However, these mechanisms introduce interaction overhead through multiple authentication steps, often degrading usability and leading to user fatigue~\cite{sasse2014great}. To mitigate this, web services have increasingly adopted alternative trust channels to verify user legitimacy, such as private links over Short Message Service (SMS).

Digital services using private links over SMS allow users to seamlessly resume authenticated sessions.
While this approach improves usability, these services offload trust to an alternative and potentially vulnerable channel. For example, private links rely on the assumption that only the intended user has access to them, which can be undermined by multiple attack vectors, such as SIM swapping~\cite{kim2022study}, cellular signaling vulnerabilities that enable SMS interception or redirection~\cite{ullah2020ss7}, malicious apps with SMS read permissions, and large-scale SMS data leaks~\cite{whittaker2019sms}.
Additionally, such URLs are often short by design. Websites may deliberately use short URLs to improve readability and fit within character limits, thereby enhancing user experience.

However, private links with predictable patterns, such as low-entropy URLs, can be exploited to make services vulnerable to large-scale account takeover or data leaks. An attacker who identifies the URL pattern can enumerate valid links through automated requests, bypassing authentication entirely without any credentials or social engineering. It is important to note that, unlike the prior case, this case requires only that the adversary know about the weakness, without requiring large-scale SMS data leaks.
Additionally, to prevent PII exposure due to accidental mishandling of SMSes, services occasionally include an authentication layer, such as a One-Time Password (OTP) or date of birth. Weaknesses in this layer of authentication can also fail under enumeration and allow unauthorized access to protected user data.

In this paper, we study private URLs to understand the threats emanating from them and the range of services affected by the threats.
Therefore, we use public SMS gateways as a lens to study this critical problem, considering ethical and non-invasive conditions~\cite{moreno2023your, gao2022demystifying, cheng2020characterizing, berenjestanaki2019exploitation, reaves2018characterizing, reaves2016sending}.
These gateways offer a lower-risk vantage point because their messages are already publicly accessible and can be observed without interacting with recipients or manipulating carrier infrastructure.
At the same time, they approximate a realistic adversarial view of SMS-delivered URLs.
Overall, this study answers two critical questions: (1) \textit{How can adversaries weaponize authentication flaws in services to infiltrate their broader user base without compromising their infrastructure?} and (2) \textit{Are such threats isolated incidents or representative of the wider ecosystem?}

Our analysis of more than 322K URLs extracted from over 33 million messages reveals concerning insights.
Using a hierarchical LLM- and expert-based PII detection and validation pipeline, we identify 701 vulnerable endpoints affecting 177 services.
Notably, 125 services use tokens with entropy below 64 bits and are therefore vulnerable to user enumeration, meaning that access to a single private URL can impact users beyond the intended recipient, including those who have never used SMS gateways or whose SMS data was never leaked.
For instance, if Alice requests insurance quotes from a provider and the URL sent to her is later obtained by an attacker, like via an SMS gateway, the privacy of other users (e.g., Bob and Eve) who previously interacted with the provider may also be compromised.
Based on publicly available user-base information for 51 of the low-entropy services, we estimate that up to 1.05 billion user records could potentially be exposed.
We further uncover 4 services that employ brute-force-able authentication mechanisms (e.g., zip code or date of birth) without rate-limiting, which are commonly used when accessing sensitive information such as medical records, raising concerns about the secure deployment of widely used technologies.
Additionally, we examine cases where SMS-delivered URLs grant capabilities beyond passive viewing. We find 20 services that expose privileged access: 14 allow modification of user records, 5 grant account-level access, and 1 allows both. These cases elevate the risk from confidentiality loss to integrity and impersonation risks.
We also identify 84 services that expose additional PII beyond the landing page.
Alarmingly, we are able to completely reconstruct 206 user profiles from SMS gateway links alone, without further enumeration of other users. This exposure could be exploited by attackers for downstream attacks, including targeted spear-phishing.
Finally, we present multiple case studies to better explain our findings, disclosing only the names of services that have been fixed and have not asked for anonymity.

Although the attacks presented in this work are widely known, our contribution contextualizes a novel attack surface that enables the accurate identification of target services without relying on blind discovery.
For example, with knowledge harvested by observing SMSes sent through public gateways, an attacker can carry out the exploitation using consumer-grade hardware and only basic to intermediate web security knowledge.
Overall, we make the following contributions:
\begin{enumerate}[topsep=0pt,itemsep=0pt,leftmargin=*]
    \item We analyze over 322K URLs from 33M+ messages collected across public SMS gateways, providing the first large-scale view of risks stemming from frictionless SMS-based user experiences.
    \item Using hierarchical LLM-based PII detection and expert validation, we uncover 701 vulnerable endpoints across 177 services, including user enumeration, brute-force-able authentication, privileged data exposure, and direct account compromise.
    \item We identify 125 services with low-entropy tokens that are potentially enumerable and confirm enumeration in 13 services under bounded testing, demonstrating how one exposed URL can extend risk to other users of the service without compromising the service infrastructure.
    \item We find 84 services that expose additional user data compared to what is visible on the landing page, amplifying privacy risks.
    \item We show that these attacks are easy to replicate using consumer-grade setups and limited expertise. Despite responsible disclosure attempts, many providers lacked reporting channels or dismissed our findings, highlighting critical gaps in industry readiness. Our responsible disclosures have been acknowledged by 18 services, 7 of which have been fixed. Only among the ones that were user-enumerable and have been fixed, we estimated \textbf{protecting at least 120 million users}.
    \item We identify a concerning trend during disclosure: many providers were skeptical, often dismissing reports as potential scams. Additionally, we found most providers lack formal mechanisms or public channels to report vulnerabilities, further hindering responsible disclosure efforts.
\end{enumerate}

\noindent \BfPara{Ethical Considerations}
This study was reviewed and approved by our institution's ethics board (IRB). We treated all collected messages, URLs, screenshots, and linked content as sensitive, even when publicly available, and ultimately deleted all raw messages, URLs, and screenshots at the end of the study.
We minimized harm through conservative crawling, local-only OCR/LLM processing, bounded URL validation, no form submissions or state-changing actions, strict data minimization, and responsible disclosure to affected providers with at least a 45-day remediation window. We also explicitly weighed residual risks. Given all the preventive measures we took, the benefits of conducting and publishing this work outweigh the residual risks.
We estimate that the resulting remediations protected at least 120 million users from an imminent threat of personal data leakage. We discuss ethics in detail in Appendix~\ref{sec:ethics}.

\section{Threat Model}
\label{sec:threat_model}

We consider an external adversary seeking unauthorized access to user PII through SMS-delivered URLs.

\BfPara{Capabilities} An adversary potentially obtains one or more SMS-delivered URLs through public SMS gateways~\cite{moreno2023your, gao2022demystifying, cheng2020characterizing, berenjestanaki2019exploitation, reaves2018characterizing, reaves2016sending}, SIM swapping~\cite{kim2022study}, cellular signaling vulnerabilities that enable SMS interception or redirection~\cite{ullah2020ss7}, malicious apps with SMS-read permissions, or large-scale SMS data leaks~\cite{whittaker2019sms}. 
We consider three adversarial scenarios: \textbf{A1} uses an ordinary web client to visit the URL, follow redirects, and inspect client-visible HTTP traffic and page content; \textbf{A2}, when a service places a secondary authentication gate after the URL, makes bounded guesses of user attributes used for authentication; and \textbf{A3} modifies the token or identifier of a valid URL when it is predictable or low-entropy, using ordinary HTTP requests to determine whether other users can be impacted.

\BfPara{Assumptions} We assume that the adversary has no legitimate credentials for the victim's account, no privileged access to the service's backend or databases, and no ability to compromise cryptographic primitives. We exclude phishing, social engineering, malware installation, carrier compromise, and denial-of-service attacks.

\BfPara{Goals} Under these capabilities and assumptions, we investigate whether exposure of an SMS-delivered URL can enable unauthorized disclosure or modification of user PII and whether the resulting impact can extend beyond the URL's intended recipient.
\section{Approach Overview}
\label{approach_overview}

\begin{figure*}[t]
    \centering
    \includegraphics[width=\textwidth]{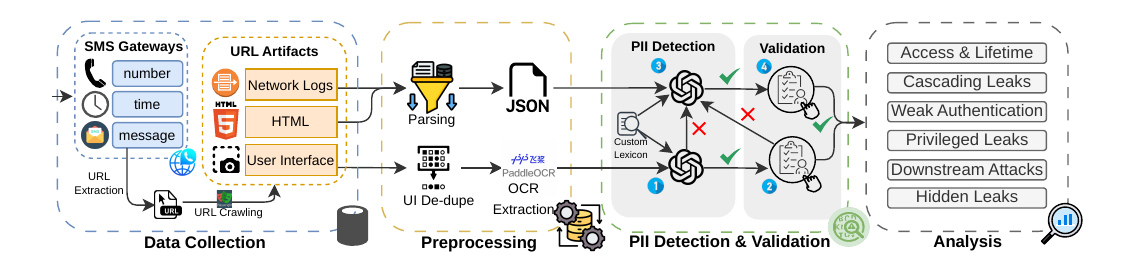}
    \caption{Our pipeline proceeds in four stages: (1) \textbf{Data Collection}: collect messages from public SMS gateways, extract URLs and crawl them using Playwright to capture Network Logs, HTML, and UI; (2) \textbf{Preprocessing}: run multilingual OCR on de-duplicated UI captures and extract JSON objects from Network Logs and HTML; (3)
    \textbf{PII Detection \& Validation}: refine the PII lexicon and use an LLM to detect PII followed by manual validation; (4) \textbf{Analysis}: analyze initial access paths and exposure lifetimes, cascading leaks from user enumeration, weak authentication, privileged data leaks, downstream user-profile risks, and hidden leaks.
    }
    \Description{Pipeline diagram with four stages: data collection from public SMS gateways and URL crawling; preprocessing of screenshots, network logs, and HTML; PII detection followed by expert validation; and analysis of access lifetime, enumeration, weak authentication, privileged access, downstream attacks, and hidden leaks.}
    \label{fig:system_design_flow}
\end{figure*}

\noindent We aim to conduct a comprehensive investigation of SMS-delivered private URLs to understand how they expose user data, what weaknesses enable these exposures, and how their risks scale across services and users. Towards this goal, we build an end-to-end pipeline to investigate the risks while being reproducible and ethical.

Figure~\ref{fig:system_design_flow} depicts our end-to-end pipeline consisting of four stages: (1) Data Collection, (2) Preprocessing, (3) PII Detection and Validation, and (4) Analysis and Results. These stages are organized in the upcoming sections as follows:

\BfPara{\underline{Data Collection and Preprocessing (\S\ref{sec:data_collection_and_preprocessing})}}
We investigate the SMS-delivered URLs through the lens of public SMS gateways, an ethical vantage point consistent with prior studies~\cite{moreno2023your, gao2022demystifying, cheng2020characterizing, berenjestanaki2019exploitation, reaves2018characterizing, reaves2016sending}. To this end, we begin by collecting messages from public SMS gateways, followed by extracting the URLs from the messages and capturing dynamic behavioral artifacts at the client-side (\ie website UI, network logs, and HTML) for unique URLs  (\S\ref{subsec:data_collection}). We then remove duplicated  UI captures and extract the embedded text using multilingual OCR on them, followed by structured extraction of JSON payloads from network logs and HTML (\S\ref{subsec:pre_processing}).

\BfPara{\underline{PII Detection and Validation (\S\ref{sec:pii_detection_and_validation})}}
We then use LLMs to flag URLs that expose PII in their artifacts (\S\ref{subsec:pii_detection}), followed by an expert validation (\S\ref{subsec:pii_validation}). To support effective detection, we construct a custom PII lexicon informed by prior literature and tailored to web-based observations. However, validating PII in UIs is considerably easier than in network logs due to the diversity in the number and size of JSON payloads.
Therefore, we design a hierarchical LLM-based PII detection and expert-annotated validation pipeline. We first detect and validate PII in UIs, followed by detecting and validating PII in payloads (for the URLs with no PII in UI).

\BfPara{\underline{Analysis and Results (\S\ref{sec:analysis_and_results})}}
Finally, we analyze the validated PII-exposing URLs to understand how exposure occurs, how it scales, and what risks it creates for users.
We begin by examining the initial access paths and exposure lifetimes of SMS-delivered URLs, focusing on whether URLs behave as bearer credentials and how long exposed links remain usable~(\S\ref{subsec:initial_access_and_lifetime}).
We then study whether a single exposed URL can lead to cascading leakage across other users of the same service by analyzing token predictability, entropy, and bounded mutation behavior~(\S\ref{subsec:user_enumeration}).
Next, we examine authentication gates that rely on user attributes such as ZIP codes, dates of birth, or surnames, and evaluate whether these gates provide meaningful resistance against guessing attempts~(\S\ref{subsec:weak_authentication}).
We further analyze the level of privilege granted by exposed URLs, distinguishing between read-only PII exposure, editable user records, and cases where URL possession grants account-level access~(\S\ref{subsec:privileged_access}).
To assess downstream risk, we examine how multiple co-located PII attributes can be combined into user profiles that may support targeted phishing, social engineering, identity theft, or financial abuse~(\S\ref{subsec:downstream_attacks}).
Finally, we look beyond the initially visible page to identify hidden data leaks caused by UI navigation and client-side overfetching, where additional PII is exposed through subsequent interactions, API responses, WebSockets, or HTML-embedded state~(\S\ref{subsec:hidden_data_leaks}).

\section{Data Collection and Preprocessing}
\label{sec:data_collection_and_preprocessing}

\noindent
SMS data collection is a crucial step to understanding the risks associated with SMS-delivered URLs.
However, the SMS data contains irrelevant messages (e.g., OTPs), duplicate landing pages (e.g., redirects to the Instagram login page), and repeated authentication pages (e.g., pharmacies asking for date of birth before showing a prescription quote).
Therefore, to limit our exposure to PIIs (further explained in section~\ref{subsec:pre_processing}), the data must be preprocessed for subsequent analysis within our pipeline.
Table \ref{tab:data_collection_and_preprocessing} describes our resultant dataset corpus, emphasized in the rest of the paper.

\begin{table}[htb]
    \centering
    \caption{Data Collection and Preprocessing}
    \scalebox{0.99}{
    \begin{tabular}{lrr}
    \toprule
        \textbf{Dataset Form} & \textbf{\#URLs} \\
        \midrule
        Messages Collected &  33,128,103 \\
        URLs Extracted & 322,949 \\
        Unique Active Final URLs & 147,251 \\
        Unique UIs & 87,080 \\
        JSON Payloads & 14,544 \\
        \bottomrule
    \end{tabular}}
    \label{tab:data_collection_and_preprocessing}
\end{table}

\subsection{Data Collection}
\label{subsec:data_collection}

\noindent We examine SMS-delivered URLs through the lens of public SMS gateways, offering an ethical perspective into the SMS ecosystem.
Our analysis begins by scraping these gateways to extract URLs and their associated dynamic attributes.

\BfPara{\underline{Collecting Messages from Public SMS Gateways}}
We compiled a list of public SMS gateways from prior work~\cite{moreno2023your, gao2022demystifying, cheng2020characterizing, berenjestanaki2019exploitation, reaves2018characterizing, reaves2016sending}. However, we realize that most of these gateways are short-lived and no longer active. We therefore used search engines and advanced search queries (Google and DuckDuckGo) to identify active alternatives.
We then filtered the ones that (1) do not require authentication, (2) are free to use, and (3) retain SMS messages for at least 24 hours. As a result, we identify 25 SMS gateways in total (complete list in appendix \ref{app:gateways}).
We developed custom scrapers using Playwright~\cite{playwrightFastReliable} for each gateway and collected all available messages by April 2024 (the date of our data collection). To minimize interference, we enabled ad-blocking~\cite{cliqzadblockerplaywright} to suppress advertisements.
Additionally, in line with the ethical considerations employed by~\cite{rautenstrauch2024s}, we introduced a two-second delay between probe requests and avoided using concurrency.
The messages collected, along with the receipt time, were stored in SQLite databases.
In total, we collected 33,128,103 messages across 30,203 unique disposable numbers.

\BfPara{\underline{Extracting URLs From Messages}} We find that gateways do not follow a consistent format when embedding URLs in messages, and some reformat message content by using line breaks, spaces, and wrapping around text. We account for the variations when extracting URLs from the messages.
As a result, we identified 485,286 messages containing at least one valid URL, out of which, we extracted 322,949 \emph{unique} URLs spanning 10,903 unique domains.
Consistent with prior work \cite{berenjestanaki2019exploitation}, the URLs are predominantly OTP links as shown by four out of the top five domains: (\texttt{t.me} (21.88\%), \texttt{ig.me} (9.07\%), \texttt{py.pl} (8.65\%), and \texttt{viber.com} (1.97\%)).
Moreover, 163,961 (50.77\%) URLs in our dataset belong to domains ranked within the Tranco top 50k~\cite{lepochat2019tranco}, highlighting that our dataset predominantly consists of prominent and widely used online services.

\BfPara{\underline{Extracting Dynamic Attributes}}
After extracting the URLs, we captured three URL attributes (UI captures, network logs, and HTML) to understand how each service handles SMS-delivered links, what authentication or authorization checks are applied, and where user data is exposed.
We iteratively visit the URLs using Playwright~\cite{playwrightFastReliable} without multithreading and with a 2-second inter-request delay to avoid imposing excessive load on remote services, following web security measurement guidelines from prior studies~\cite{rautenstrauch2024s}.
Al-Fannah \etal~\cite{al2018beyond} reported an average full-page load time of 19s across 10K websites and recommended that crawlers should not time out before at least 20s. Therefore, we wait for the \texttt{networkidle} state, with a 20\,s timeout, before proceeding to the next URL to account for slow-loading pages and delayed network activity.
We removed duplicate URLs previously, but unique URLs can lead to the same final URL due to redirects. Therefore, we capture the full-page UI screenshot (.png), network logs (.har), and HTML snapshot (.html) only for unique final URLs to prevent redundant data collection.

Overall, we captured the attributes of 147,251 (45.60\%) URLs from the initial set of 322,949 URLs. By collecting artifacts only for unique final redirected URLs, we stopped ourselves from performing recurring collection for 83,419 (25.83\%) URLs, and 92,279 (28.57\%) were inaccessible due to network timeouts or connection errors.
One reason for the high inaccessibility rate is that we visited these URLs at least a year after the messages were received because we wanted to focus on long-lived PII exposing URLs and avoid ephemeral and one-time access links, including OTP flows.

\subsection{Preprocessing}
\label{subsec:pre_processing}

\noindent Data collection provides three raw artifacts per URL: UI captures, network logs (HAR), and HTML. However, some artifacts may correspond to short-lived URLs, leading to error pages such as \textit{404 -- Not Found} or duplicates when different URLs produce the same website view. Therefore, preprocessing transforms the raw UI, HAR, and HTML artifacts into a deduplicated set of analyzable textual and structured representations suitable for scalable PII detection and validation.
Preprocessing is performed at two levels: (1) UI, and (2) Network and HTML. The UI stage filters duplicate views caused by redirects or error pages and extracts text from captures using OCR. The Network and HTML stage focuses on extracting JSON payloads from network logs and HTML content. The output of this module is a cleaned, deduplicated dataset of UI text and JSON data, which facilitates the PII detection and Validation step (\S~\ref{sec:pii_detection_and_validation}).

\BfPara{\underline{UI Pre-processing}}
Web services often generate URLs that redirect users to generic or repeated pages, such as Instagram login screens before displaying content, pharmacy date-of-birth prompts before showing prescription details, or shared error pages for expired or broken links.
To address this, we only retain unique UI captures by removing byte-identical screenshots of the rendered URLs.
We then locally run multilingual OCR on each retained UI to extract textual data from it.
While HTML or DOM-based extraction is possible, robustly extracting only rendered, visible text from them at scale is challenging. First, an HTML-only analysis would miss dynamic content that is loaded and rendered \textit{dynamically}. Second, DOM-based analysis would miss rendered content outside the DOM (e.g., images, PDFs, canvases), which OCR reliably captures. OCR also captures text rendered outside standard DOM text nodes (e.g., images/canvas). We complement this UI-visible view with network-intercepted data (i.e., HTML, JSON) to analyze non-visible leaks.

\BfPara{\underline{De-duplication}}
We found that multiple URLs resolved to visually identical pages (e.g., shared error handlers).
Our initial approach explored perceptual/structural similarity methods (e.g., SSIM, pHash, feature matching) because they can group pages that are visually similar despite minor rendering differences. However, we realized that they can over-merge pages where ``\textit{small}'' visual differences encode sensitive identifiers (e.g., names, addresses, order IDs).
Considering the loss of critical details, we resort to only removing byte-level screenshots by computing MD5 checksums of all UI captures and retaining only the first occurrence of each unique hash.
As a result, we identified 87,080 (60.37\%) unique UIs (3,958 domains) out of the 147,251.
It is essential to note that this method does not discount near-duplicate UIs (e.g., distinct OTP codes). Therefore, our resultant dataset of unique UIs is still dominated by OTP-verification destinations, mainly \texttt{t.me} (72.69\%) and \texttt{viber.com} (4.92\%). The remaining entries in the top ten domains are global or custom URL shorteners that redirect to heterogeneous pages. In the deduplicated dataset corpus, we found that 76,524 (87.88\%) URLs were ranked within the top 50k according to the Tranco list \cite{lepochat2019tranco}.

\BfPara{\underline{OCR Extraction}}
We take an LLM-based approach to identify PIIs in the UI. This necessitates extracting text from the UI captures, formatted as images. A reasonable option is utilizing the HTML to extract the text instead of extracting text from images.
However, HTML-based text extraction has its own challenges, such as missing the contents rendered outside the DOM (we discuss this in detail in this subsection above under ``UI Pre-processing'').
Additionally, we use a text input for LLMs instead of passing UI captures directly to a vision or multimodal model because the target signals (PIIs) in our study are primarily textual identifiers. Furthermore, in the vision or multimodal model design, the LLM would need to both extract text from the screenshots and detect PIIs in that text. However, our current design relies on tools specialized to extract text from the images.
Additionally, public SMS gateways are accessible around the globe, making the messages linguistically diverse and the webpages heterogeneous~\cite{reaves2016detecting}.
Therefore, we employ a robust multilingual OCR pipeline based on PaddleOCR 3.2.0 (PP-OCRv5, Universal Scene Text Recognition) \cite{cui2025paddleocr30technicalreport} to extract embedded text from UI captures. In our deployment, PaddleOCR runs inference locally, ensuring that potential PII is not uploaded to external servers.

\BfPara{\underline{Network and HTML Preprocessing}}
Websites typically make API or WebSocket requests to render the UI and can have information embedded in the HTML that is invisible in the UI~\cite{pan2024edefuzz,owaspAPI32019Excessive}.
Therefore, to conduct a comprehensive investigation of the client-side behavior, we focus below on extracting JSON data objects embedded in the HTML, as well as JSON responses from APIs and WebSockets. We limit our analysis to JSON payloads because they commonly contain user data~\cite{owaspAPI52023Broken}, while substantially reducing the volume of data to be processed and validated.
Together, the extracted JSON objects and OCR-derived UI text provide a normalized representation of visible and client-side data, allowing subsequent PII detection to be performed more efficiently by reducing redundant data and limiting analysis to relevant textual and structured content.
While parsing the network logs (stored as HAR files), we select response bodies based on Content-Type. If a body is JSON, we parse it directly; if it is JSONP, we unwrap the function call and parse the payload. Similarly, if a body is HTML, we extract embedded JSON from (1) \texttt{<script>} tags whose \texttt{type} is \texttt{application/ld+json}, \\  \texttt{application/json}, or \texttt{application/manifest+json}, (2) the widely used \texttt{\_\_NEXT\_DATA\_\_} blob, and (3) inline assignments whose right-hand side is strict JSON (e.g., \texttt{window.* = \{...\};}). We then perform JSON extraction over the saved HTML files. After canonicalizing and de-duplicating extracted JSONs using SHA-256 hash, we extracted 14,544 JSON payloads from 87,080 unique URLs.

\section{PII Detection and Validation}
\label{sec:pii_detection_and_validation}

\noindent The preprocessing module provides us with the text in the UIs and normalized JSON snippets (from Network Logs/HTML).
Since preprocessing converts raw artifacts into unique and structured representations, the data is now suitable for LLM-based detection of potential PII exposures, followed by expert validation.
Therefore, this section describes the PII detection and validation module.

Prior work has largely relied on the implementation of rule-based matching against raw strings~\cite{190900,he2024maginot,senol2022leaky}; however, this technique produces brittle results when multiple languages, templates, and layouts are involved.
On the other hand, recent studies have explored using LLMs to identify PII~\cite{ren2016recon,rajgarhia2025evaluation,cheng2025effective} and the associated risks of hallucinations, leading to inconsistent labels.
We therefore design a pipeline where an LLM flags potential PII exposures, and expert reviewers validate PII-exposing URLs.

\begin{table}[tb]
  \caption{Custom PII lexicon creation. We retained 66 PII types from 287 policy-derived terms}
  \centering
  \scalebox{0.92}{
    \begin{tabular}{llr}
      \toprule
      \textbf{Removal Reason}              & \textbf{Examples}                    & \textbf{\#Keywords} \\
      \midrule
      Repetition                           & \emph{(location, geolocation)}       & 86 \\
      Generic/Vague                    & \emph{(personal data, physical)}     & 75 \\
      Not Likely in Text                   & \emph{(visual, thermal)}             & 40 \\
      Not Likely in Webpage                & \emph{(device identifier, beacons)}  & 18 \\
      OTP or Login Code                    & \emph{(access code, security code)}  & 2  \\
      \midrule
      \textbf{Total Removed}               & \emph{ }                             & \textbf{221}  \\
      \bottomrule
    \end{tabular}}
  \label{tab:pii_lexicon_reduction}
\end{table}

\BfPara{\underline{Hierarchical Detection and Validation}}
The preprocessed artifacts contain two complementary views of each URL: UI text that approximates what a link holder can directly observe, and JSON payloads that may expose additional client-side data not rendered in the browser. We therefore use a hierarchical strategy to prioritize the lower-volume, more interpretable UI view first, and then analyze JSON payloads only where UI analysis does not already confirm an exposure. This ordering reduces redundant LLM processing and reviewer effort and improves validation reliability because PII in visible UI is easier for reviewers to inspect in context. After the UI-based validation, we run the JSON-based detection LLM on data that (1) the UI-based LLM flagged as non-PII or (2) was discarded during the manual UI-based validation. Finally, we also validated the JSON payloads flagged by LLM.

\BfPara{\underline{PII Taxonomy}}
Before we attempt to use LLMs to identify PII, it is essential that we ground our definition of PII on the foundations of the extensive research in this space.
While the privacy domain has resulted in a multifaceted definition of PII, covering various use cases (e.g., biometric information, health records, advertisement interactions), it is important that we reduce the definitions to a narrower web domain, enabling efficient LLM adoption by saving on token usage.
We utilize the taxonomy created by Lovato et al.~\cite{lovato2023more} as a baseline because they provide a comprehensive and privacy policy-derived lexicon of PII types.
For the 287 keywords in the original lexicon, we removed 221 based on the following filtering strategy:
    (1) \texttt{repetition:} When a PII type is similar to another, keep only the most clear and/or standard term,
    (2) \texttt{too generic/vague:} When a PII type is overly generic or vague and does not meaningfully represent identifiable information,
    (3) \texttt{not likely in text:} When a PII type can not be directly identified from text,
    (4) \texttt{not likely in webpage:} When a PII type is not likely to be present in or verified just from the  website UI or JSON payloads, or
    (5) \texttt{otp or login code:} When a PII type is related to OTPs or Login Codes.
Consequently, we retain 66 unique PII types relevant to webpage UI and JSON artifacts (see Appendix~\S\ref{app:pii_taxonomy}). The most frequent removal cause was repetition (86 terms, 38.91\%), followed by overly generic or vague descriptors (75 terms, 33.94\%). Table~\ref{tab:pii_lexicon_reduction} summarizes the removal reasons for the 221 PII types in the original PII lexicon.

\subsection{PII Detection}
\label{subsec:pii_detection}

This subsection focuses on detecting PII in the textual data from the preprocessing stage. However, given the volume and complexity (of OCR text \& JSON payloads), we need a scalable detection pipeline.
Traditional methods like rigid pattern matching miss context and provide unreliable results.
However, LLMs have been shown to be effective for privacy-relevant text analysis because they can capture semantic context beyond rigid patterns and support flexibility across heterogeneous inputs. For instance, Liu et al.~\cite{liu2025evaluating} show that LLM-based PII extraction outperforms expressions, keyword search, and entity detection.

A rule-based alternative would require maintaining hand-crafted detection heuristics for 66 PII types across multilingual text, heterogeneous webpage templates, and structurally diverse JSON payloads. More importantly, our task required contextual understanding. For example, a name may identify the user, appear as public company information, or occur as unrelated content on the page. We therefore use the LLM as a scalable candidate-detection stage followed by manual validation.

We constrain our LLM to classify PII based on our improved set of PII identifiers with a strict output schema.
For each input, the LLM first determines whether PII is present and, if so, labels the exposure with one or more taxonomy-defined PII types, such as first name, phone number, or email address.
We apply a zero-shot LLM pass over OCR text for UI-based detection and JSON objects for the JSON-based detection.
Inference runs locally via LM Studio~\cite{LMStudio2025} using the open-source \texttt{gpt-oss-120b} model, which has been shown to match the performance of GPT-4o-mini~\cite{agarwal2025gpt}. This setup ensures that no content is sent to third-party services and that all data remains within the analysis environment.
We construct a \emph{system prompt} that sets high reasoning and enforces an exact output schema, and a \emph{user prompt} that includes the reduced PII lexicon and the OCR text or JSON payload.

\subsection{PII Validation}
\label{subsec:pii_validation}
LLMs are known to hallucinate, making it difficult to formulate conclusive remarks. Therefore, it is essential that we include a human validation step to mitigate false positives and label drift~\cite{ji2023towards}.
For us, validation ensures that data leaks correspond to real, user-linked PII and allows us to categorize their type.

To ensure label correctness and reduce individual bias, two expert annotators independently review all LLM-flagged items, first for UI and then for JSON payloads.
Reviewers label each item \emph{true positive} (PII present) or \emph{false positive} (no PII) and correct the categories when needed. Reviewers applied the following criteria when labeling something as \emph{false positive}:
    (1) \texttt{PII-asking page:} asking to provide PII in forms,
    (2) \texttt{sample/demo content:} prefilled with sample content like example.com, ads with sample content, or testing dummy data in dev sites,
    (3) \texttt{public or company information:} website's phone number, email, address or anything that is intended to be public like names on social media, reviews with users' names, or disposable phone numbers,
    (4) \texttt{keyword match:} Exact or similar keyword match for PII like Name for name or SSN for social security, or
    (5) \texttt{misclassification:} it is a wrong match, mostly due to partial info like (user*****@gmail.com).

After labeling by the two expert annotators, we resolved disagreements by discussion and joint review. In case of no agreement after discussion, \textbf{we label it as a false positive to avoid overestimation}. After conflict resolution, we obtain the final dataset containing labels and categories for further analysis. We use the following methodology for UI and JSON-payload validation.

\noindent \BfPara{UI Validation} We opened the UI capture for each of the flagged URLs and looked for visible user PII.
We fix the PII types if the LLM over-labeled or under-labeled any true positive. If the LLM flagged a UI as having PII but we do not see any user PII (due to any of the reasons described earlier), we label it as a false positive.

\noindent \BfPara{JSON Payload Validation}
Similar to the UI validation, we begin by analyzing the raw LLM output. However, considering the size and complexity of the payloads, we specifically focus on the identified PII.
We open the JSON snippets for the URL and search for the PII; otherwise, we flag the sample as false.
We do so because of the time required for the comprehensive analysis of every payload.
We skim through the parts around that keyword to understand the context. For example, Fig.~\ref{fig:stickermule} (in appendix~\ref{app:case_studies_figs}) shows that the exposed email appears within the \texttt{order} object and alongside order-specific fields. This context confirms that the email is associated with the user’s order rather than belonging to a developer, company employee, or unrelated log entry.
Based on this information, we only label something as a true positive if we are certain that the exposed PII is the user's private information.

\subsection{Detection and Validation Results}
\label{subsec:hierarchical_detection_and_validation}

\BfPara{\underline{UI Results}}
Our LLM detection over OCR text for all 87,080 unique UIs flagged 1,916 (2.20\%) candidates for manual validation. Reviewer one labeled the 538 \emph{true positives} while reviewer two labeled 677 \emph{true positives}. After conflict resolution, we confirmed 553 \emph{true positives} (PII present) and labeled 1,363 as \emph{false positives}.
The common cause of discrepancies in this step was that one reviewer treated the exposure of the SMS gateway number as user PII; however, a consensus was reached that it should not be labeled as PII, as this number is already compromised or known to the adversary.

\BfPara{\underline{JSON Payloads Results}}
We found that 14,544 URLs had JSON payloads in their network logs or HTML, 440 of which were marked as having PII during UI validation.
Our LLM detection over JSON payloads for the 14,104 payloads flagged 1,958 (13.88\%) candidates for manual validation. Reviewer one labeled the 163 \emph{true positives} while reviewer two labeled 161 \emph{true positives}. After conflict resolution, we confirmed 148 \emph{true positives} (PII present) and labeled 1,810 as \emph{false positives}. The main source of false positives at this stage was the LLM’s tendency to flag public information, such as company contact details, as PII, even though the prompt explicitly instructed the model to ignore such content.
During this step, the reviewers had comparatively fewer discrepancies due to experience from the UI pass. The only conflicts arose when reviewers could not tell from the JSON payloads whether the PII under investigation was relevant to the user. For example, we did not label a non-company email address as user PII unless nearby JSON fields provided concrete evidence linking it to the user.

\begin{table}[t]
    \centering
    \caption{PII Detection and Validation Metadata}
    \scalebox{0.99}{
    \begin{tabular}{lrr}
    \toprule
        \textbf{Dataset Form} & \textbf{\#URLs} & \textbf{\#Services} \\
        \midrule
        Initial Set of URLs &  87,080 &  3,958 \\
        PII Detected in UIs & 1,916 & 819 \\
        PII Validated in UIs & 553 &  121 \\
        PII Detected in JSON payloads & 1,958 & 844 \\
        PII Validated in JSON payloads & 148 & 67 \\
        \bottomrule
    \end{tabular}}
    \label{tab:detectNValid}
\end{table}

\BfPara{\underline{Validated PII Exposing URLs}}
In total, we validated 701 URL endpoints across 177 services. Table \ref{tab:detectNValid} summarizes the results from detection and validation. The high false-positive rate primarily reflects the difficulty of distinguishing user-linked PII from syntactically similar but non-user information. For example, many flagged values were genuine names, email addresses, or postal addresses, but belonged to developers, companies, or other public entities rather than the user associated with the URL.
During manual validation, we confirmed that none of the 177 affected services were scams. We also found that 55 of the services appear in the tranco top 1M list, including 14 ranked within the top 50K.

\section{Analysis and Results}
\label{sec:analysis_and_results}

\begin{table*}[thb]
    \centering
    \caption{PII analysis results and root--cause mapping with OWASP/API Top-10 and CWE}
    \scalebox{0.93}{
    \begin{tabular}{llll}
    \toprule
    \textbf{Analysis Category} & \textbf{OWASP / API Top-10} & \textbf{CWE} & \textbf{Key Result} \\
    \midrule
    \textbf{Access \& Lifetime~\S\ref{subsec:initial_access_and_lifetime}}
      & A01:2025; API2:2023
      & CWE-284; CWE-287; CWE-862
      & Token (Server: 79; API: 92; Socket: 1), URL: 5 \\
    \addlinespace[3pt]
    \textbf{Cascading Leaks~\S\ref{subsec:user_enumeration}}
      & A01:2025; A04:2025; API1:2023
      & CWE-331; CWE-307; CWE-425
      & Low Entropy: 125, Enumerable: 13 \\
    \addlinespace[3pt]
    \textbf{Weak Authentication~\S\ref{subsec:weak_authentication}}
      & A06:2025; A07:2025
      & CWE-287; CWE-307
      & Weak Authentication: 4 \\
    \addlinespace[3pt]
    \textbf{Privileged Leaks~\S\ref{subsec:privileged_access}}
      & A01:2025; A06:2025; A07:2025
      & CWE-359; CWE-862; CWE-613
      & Editable: 14, Account Access: 5, Both: 1 \\
    \addlinespace[3pt]
    \textbf{Downstream Attacks~\S\ref{subsec:downstream_attacks}}
      & A01:2025
      & CWE-200; CWE-359
      & User Profiles: 206 \\
    \addlinespace[3pt]
    \textbf{Hidden Leaks~\S\ref{subsec:hidden_data_leaks}}
      & API1:2023; A06:2025
      & CWE-201; CWE-425; CWE-862
      & Overfetching: 76, UI Interactions: 8 \\
    \bottomrule
    \end{tabular}}
    \label{tab:analysis_results}
\end{table*}

This section describes the analyses of the validated PII leaks and our findings.
Specifically, we examine the initial access paths and exposure lifetimes of leaked URLs (\S\ref{subsec:initial_access_and_lifetime}), cascading leakage through user enumeration (\S\ref{subsec:user_enumeration}), weak authentication secrets (\S\ref{subsec:weak_authentication}), privileged account access (\S\ref{subsec:privileged_access}), potential downstream attacks (\S\ref{subsec:downstream_attacks}), and hidden data leaks caused by client-side over-fetching due to improper server-side filtering (\S\ref{subsec:hidden_data_leaks}). Table~\ref{tab:analysis_results} summarizes the analyses, OWASP/CWE mappings, and key results across \S\ref{subsec:initial_access_and_lifetime}–\S\ref{subsec:hidden_data_leaks}.

Whenever necessary, we explain our findings and emphasize the root cause with an example vulnerable service. It is essential to note that we conduct the below analysis under strict ethical guidelines as detailed in~\autoref{sec:ethics}. Additionally, we only name the service if they have already fixed the vulnerability.

\subsection{Initial Access and Lifetime of URLs}
\label{subsec:initial_access_and_lifetime}

SMS-delivered private URLs are intended to provide temporary, low-friction access, but they become risky when possession of the link is sufficient to retrieve user PII.
This risk depends on both how the service makes the private webpage accessible to the non-intended user (aka adversary) and how long the URL remains active. The exposure path reveals the root cause of the leak, while the URL lifetime determines how long the link can be logged, scraped, forwarded, leaked, or rediscovered.

\BfPara{\underline{Methodology}}
We open the URL with DevTools enabled to observe redirect links, network logs, and headers to pinpoint the initial source of PII exposure. We identify the exposed PII using the search feature under the Network tab in DevTools. The text search gives us item(s) that has the PII and then we manually step through them in reverse chronological order to locate the first item that contained PII (or the item that requests it). During the process, we check the ``Headers'' and ``Response'' tabs to analyze each request/response chain. In case of tokenized URLs, we examine API calls and WebSocket messages for token usage. If a token is used in the request header but no API or socket activity is visible, we classify the exposure as \textbf{server-rendered}. If we find the PII in the original URL or the redirect chain headers, we mark it as \textbf{URL-embedded PII}.
We also measure exposure lifetime, as having long or permanent access URLs significantly increases the probability of exploitation by an attacker. For this reason, sensitive URLs must be only accessible for a short and defined time~\cite{owasp-session-management}.
We measure the exposure time as the difference between the SMS delivery date and the URL crawling date \textbf{(July 26, 2025)}. Because all validated URLs were still active on the crawl date, the exposure time we report is a lower bound on their active duration. It is essential to remember that even if a domain were re-registered by a different entity between the time the message was sent and our crawl, it is extremely unlikely that the same access token would be reused. To further understand the PII exposure behavior, we categorize the services (domains) by their associated industries. Similar to prior work~\cite{mvpnalyzer}, we utilize the VirusTotal Forcepoint~\cite{forcepoint} classification engine to obtain the industry for each domain. In total, the Forcepoint engine was able to classify 118/150 domains whose data was extractable using VirusTotal API; the list of categories is presented in Table~\ref{tab:url_categories}.

\begin{figure}[th]
    \centering
    \includegraphics[width=0.95\linewidth]{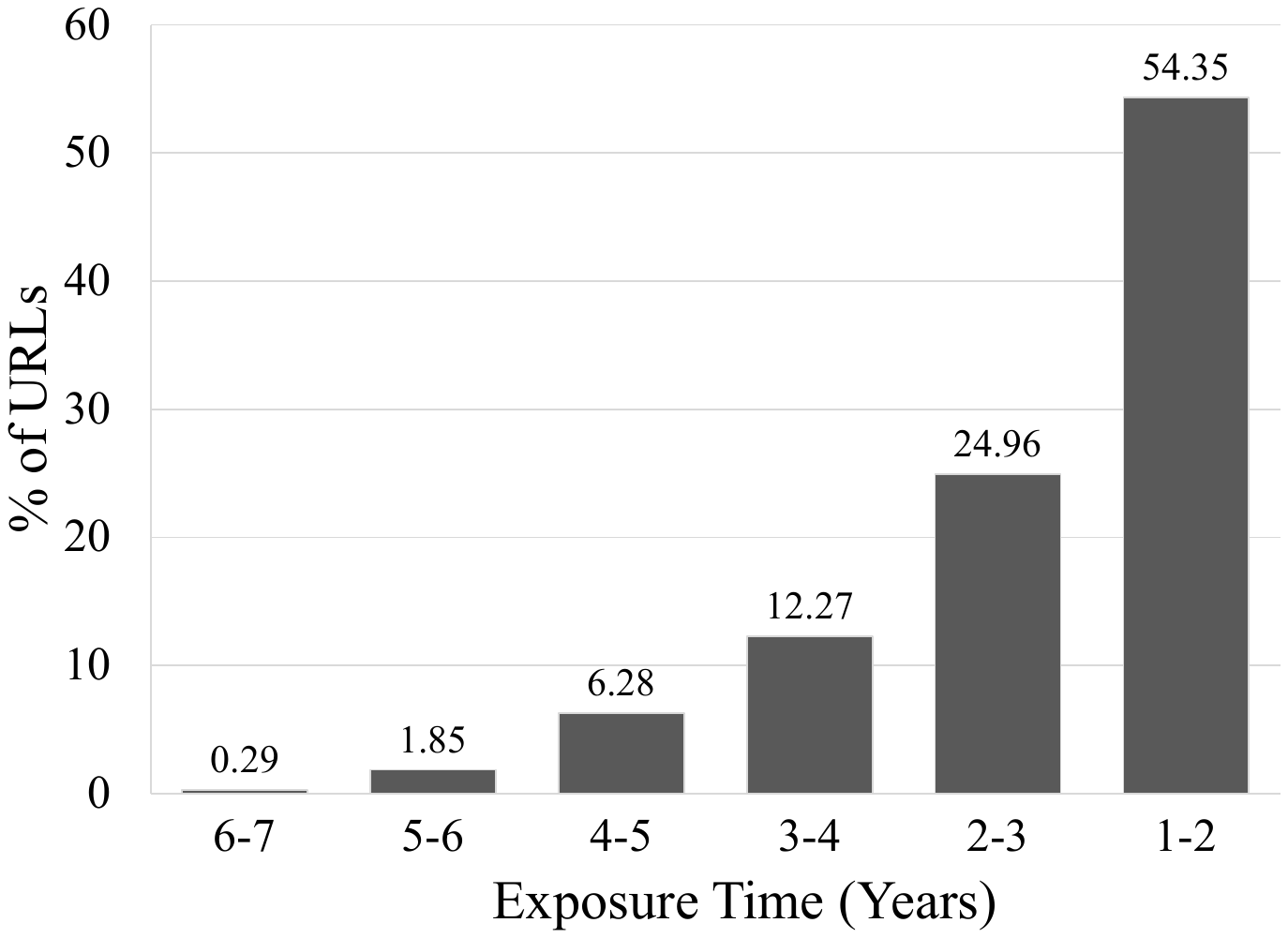}
    \caption{PII-Exposing URLs by Exposure Time (n=701)}
    \Description{Bar chart of the lower-bound active lifetime of 701 PII-exposing URLs. The largest group, 54.35 percent, remained active for one to two years; 24.96 percent remained active for two to three years, with smaller fractions remaining active for up to seven years.}
    \label{fig:entries}
\end{figure}

\BfPara{\underline{Results}}
Tokenized bearer links were the dominant root cause, totaling 172/177 (97.18\%) services. Of these,  92 (53.49\%) called an API, 79 (45.93\%) were server-rendered, and 1 (0.58\%) used a WebSocket. In the remaining 5/177 (2.82\%) cases, PII appeared directly in an original URL or redirect chain.
As shown in Figure~\ref{fig:entries}, we observe that 54.35\% of PII-exposing URLs were sent within a period of 1 to 2 years, and the remaining 45.65\% were at least two years old. The drop in exposure time from 1-2 to 2-3 years suggests that some URLs may expire after 2 years, but we believe it is because of dataset bias as only a handful of public SMS gateways store SMS data for longer than 2 years. Still, we see PII-exposing URLs leading back to 2019. This indicates the presence of long-lived URLs, and when combined with PII exposure, it leaves the user's PII at a higher risk of attack. Additionally, we find that the most common categories are ``business'' and ``IT'' with 27 and 23 services, respectively. Surprisingly, \textbf{19 services belong to financial services}, contrasting with the stricter expected levels of security regulations in the financial industry~\cite{arcticwolf_financial_regulations}. Table~\ref{tab:url_categories} shows popular categories for 105 of the 118 service categories classified by Forcepoint.
Various security regulations such as GDPR~\cite{gdpr} and guidelines like NIST SP 800-122~\cite{nist800-122} advocate for secure data retention policies. Although NIST clearly states that PII should not be stored for long periods, they lack specifying explicit time-frame to enforce provider compliance and allow PII retention when it is ``strictly" necessary.

\begin{table}[H]
  \caption{Industry categorization of service providers (domains) using Forcepoint classification engine}
  \centering
  \scalebox{0.99}{
    \begin{tabular}{lr|lr}
      \toprule
      \textbf{Industry} & \textbf{\#Services} & \textbf{Industry} & \textbf{\#Services} \\
      \midrule
      Business & 27 & Job Search & 4 \\

      \cellcolor{lightred} IT
        & \cellcolor{lightred} 23
        & Real Estate & 4 \\

      \cellcolor{lightred} Finance
        & \cellcolor{lightred} 19
        & Educational & 3 \\

      Shopping & 9 & Business App. & 2 \\
      Vehicles & 6 & Restaurants & 2 \\
      Health & 4 & Media & 2 \\
      \bottomrule
    \end{tabular}
  }
  \label{tab:url_categories}
\end{table}

\BfPara{\underline{Impact}}
This ``URL-as-credential'' antipattern amplifies risk in SMS workflows where messages are routinely logged or intercepted. Operationally, these patterns treat the URL itself as the authenticator. The information about the source of the data leak helps guide further analysis and provide informed recommendations to the security community.
We also found that approximately 46\% of URLs remained active after 2 years, significantly increasing the risk of attacks for these users.
We argue that URLs exposing or granting access to PII should have short, explicitly defined lifetimes based on the sensitivity of the exposed data and the expected user workflow, and that access to those records should remain restricted behind strong authentication gates.
These gaps are particularly concerning for critical sectors such as ``IT'' and ``Finance'', which are considered critical infrastructure~\cite{cisaCriticalInfrastructure} (highlighted in Table~\ref{tab:url_categories}). URLs from ``Finance'' mostly led to pre-filled loan applications, which could reveal users' financial status~\cite{pets14-https-tracking}, making them vulnerable targets for scammers.

\takeaway{Tokenized bearer links account for $\approx$97\% of observed PII exposures, and 45.65\% of validated PII-exposing URLs remained active for more than two years. SMS-delivered private URLs should not act as standalone authenticators: they should be short-lived, one-time use where possible, protected by server-side authorization checks, and governed by clear industry-specific lifetime limits.}

\subsection{Cascading Leaks from User Enumeration}
\label{subsec:user_enumeration}

\noindent A leaked SMS-delivered URL can expose the intended recipient's data. However, predictable identifiers within URLs can expand that harm to other users of the service. In other words, if tokens are low-entropy or derived from predictable sequential identifiers, an adversary who has access to only one valid URL would be able to guess neighboring URLs and access records of the entire user base.

\BfPara{\underline{Methodology}}
Due to the higher percentage of observer URLs being tokenized bearer URLs, we analyze their token entropy to assess whether these tokens are enumerable. We follow the theoretical foundations laid by MITRE's Common Weakness Enumeration, specifically CWE-331: Insufficient Entropy \cite{CWE331_InsufficientEntropy}, and OWASP \cite{OWASP_InsufficientSessionIDLength} guidelines. For every token identified during root-cause analysis, we estimated its entropy as 
$H = \log_2(|\mathcal{A}|^L)$
, where $L$ is the token length and $|\mathcal{A}|$ is the effective alphabet size. We experimentally verified whether case sensitivity contributed to entropy by flipping the first alphabetic character in the URL and if the response was the same as the UI of the original URL, the alphabet's possible values were reduced from 52 to 26. If the token only contained hex characters (0-9 and A-F), we assumed it is a hex token (16 possible values per character) and not alphanumeric character (36 possible values per character).  Tokens with $H < 64$ bits were flagged as potentially enumerable. Further, we manually attempted up to 10 token mutations for every tokenized URL and stopped at the first hit. We then classified a domain as enumerable if a modified URL yielded another user’s PII.

\BfPara{\underline{Results}}
Among the 172 tokenized URL services (recall, we select one representative URL per final domain), 125 (72.67\%) services have a token with entropy below the 64-bit threshold and were flagged as potentially enumerable. During the 10 manual token mutations, we obtained 13 successful hits (10.40\%). In these cases, we confirmed only the existence of exposure and did not store screenshots, network logs, or PII due to ethical concerns, since these URLs were not directly retrieved from public SMS gateways.
Apart from entropy, enumeration also depends on the density of valid users within the token space. However, our study does not estimate this density or attempt to confirm large-scale enumerability, because doing so would require automated token mutation and requests at a scale that would be ethically inappropriate.

\noindent \textit{Example (Anonymous Transport Company):}
It is an auto transport company working across 49 states. We find that quotes assigned to users have a unique token assigned, formatted as \texttt{123456-AB}. However, the pattern was easy to identify: changing the token (which is also the order number) to \texttt{12345\textbf{7}-AB} resulted in access to the next user's quote. Further, the URL in our dataset had the order number around 650K (the 123456 in the example URL), but checking an order number in the range of around 300K resulted in another user's quote. Therefore, it shows that linear scanning may reveal at least 650K users' PII. Also, this issue was not unique to this service: four other services among the 13 confirmed enumerable cases made similar design choices.

\BfPara{\underline{Impact}}
Low-entropy or predictable tokens turn a single leaked URL into a horizontal compromise. Once an adversary learns the token format, they can probe adjacent or otherwise guessable identifiers using ordinary HTTP requests, potentially discovering additional users' records without compromising the provider's infrastructure. The risk increases when many valid records occupy a small or sequential token space: each successful guess requires less time and effort due to the saturated token space. Consequently, even users whose SMS messages were never exposed can be affected if the service uses predictable SMS-link identifiers.

\takeaway{We found that 125 services use tokens with estimated entropy below 64 bits and are therefore potentially enumerable. For 13 of those services, bounded testing confirmed access to another user's data within the first 10 mutations. Service providers should use cryptographically random, URL-safe tokens with at least 64 bits of entropy.}

\subsection{Weak Authentication Secrets}
\label{subsec:weak_authentication}

\noindent
For a majority (97.18\%) of the services vulnerable to data leaks, we find that the adversary can directly get access to user data. However, we also find that services in our dataset request authentication secrets, such as an OTP or date of birth, before exposing user data. As intended, we assume only users know this secret and consider social engineering-based attacks out of scope. It is also expected that services implement a limit on how many times a user is allowed to guess the authentication secret. Not doing so creates an avenue for typical brute force attacks. Below, we describe our approach to uncovering these potential attacks and present our findings.

\BfPara{\underline{Methodology}}
For pages requiring authentication before granting access, we check whether the required value appears in our SMS corpus for that number or is trivially guessable (e.g., ZIP code). We also inspect network logs for clues that could reduce the cost of enumeration for an attacker. When encountering URLs with such authentication constraints, we perform up to ten automated attempts, stopping immediately upon a successful match, exhaustion of attempts, or any indication that further attempts are blocked. During this process, we observe server responses and timing behaviors to identify lockouts, CAPTCHAs, or escalating delays.

\BfPara{\underline{Results}}
Among the 177 services, five services show limited PII in the initial view but require authentication to access more privileged information such as health records. We find that four of these services implemented weak authentication parameters. These services were also vulnerable to weak or absent rate-limiting as we were not blocked during the 10 automated attempts.

\noindent \textit{Example (Anonymous Financial Company):}
It is a debt purchase company that has helped over 2 million customers. As a well-known financial company, it protects users' private information by requiring two authentication secrets before showing sensitive data. Unlike the other two of the four cases in our dataset, which required one secret, this appears more secure. However, by searching the SMS data for the same number using the given first name as a keyword, we were able to find one of the authentication secrets, \ie the surname. Consequently, we were left with only one enumerable secret (date of birth) which was not protected by rate limiting.

\BfPara{\underline{Impact}}
Absence of throttling allows the authentication to collapse under trivial enumeration and become guessable credentials, turning a leaked link from a read-only privacy breach into a pathway for account takeover or broader PII exposure. Additionally, when combined with predictable tokens, this can create a false sense of security while still enabling a service-wide cascading data leak.

\takeaway{We identify 4 services with enumerable authentication. This underscores the need for strict limits on the number of authentication attempts to prevent easy guessing of secrets such as last name, ZIP code, or date of birth.}

\subsection{Privileged Data Leaks}
\label{subsec:privileged_access}

In the previous subsections, we describe the risks of private data being exposed to an adversary. However, the impact can be more severe when possession of the URL grants privileged access. This could be in form of edit access to user records and/or logged-in access to user's account.
Such exposures escalate the impact from confidentiality loss to integrity risk.

\BfPara{\underline{Methodology}}
To determine the exposure level, we analyze only the first loaded page's UI to check if the exposed PII is editable or read-only from the adversary's perspective. If a URL lands the adversary on a page containing PII that they can edit, we consider it as \textit{editable}. Otherwise, if the PII is not editable or available in the JSON payloads, the exposure is termed as \textit{read-only}. We also confirm whether the endpoint is in a logged-in state by checking whether the rendered view has authenticated-only navigation elements (e.g., “My Account,” “Billing,” “Logout”) or personalized headers (“Welcome, \texttt{<name>}”), or pages that enumerate private resources (applications, orders, messages).

\BfPara{\underline{Results}}
The majority of cases (162/177) were read-only data spread across UI (103/162) and JSON payloads (59/162). The remaining (15/177) editable data instances were pre-filled forms in the UI. Beyond editable forms, we identified six services that granted account access based solely on possession of the URL. One of the account-access cases overlapped with the editable cases. Therefore, we found 20 unique services with privileged access.

\noindent \textit{Example (iHire):}
It is an employment platform that brings candidates and employers together in 57 industry-focused communities such as iHireDental, iHireConstruction, and iHireChefs. They are used by 103K organizations (\eg Walt Disney World, Marriott, Lowe’s) and have ``3.45+ million active members in the last 30 days'' \cite{ihireIndustryFocusedBoards}, implying broad impact. They use SMS and email to send job application updates, including links with short tokens (five-character alphanumeric). These links redirect anyone to a page with ``Welcome back, FirstName'' and a ``Continue'' button underneath it (refer to appendix Fig.~\ref{fig:ihire}). Clicking on the button redirects to the logged-in profile with edit (and read) access to resume, contact details, address, job applications, and email \& password for the account. This was also the only account-access case that overlapped with the editable cases, enabling account takeover by allowing the password and email address associated with the account to be changed without any authentication.
It is also concerning because low token entropy, combined with a large user base, resulted in a saturated token space. Consequently, we were able to access another user’s account only after two token mutations, which once again reiterates the
cascading effect described in \S\ref{subsec:user_enumeration}.

\BfPara{\underline{Impact}}
These flows collapse the boundary between ``viewing a page'' and ``acting as the user'' of the service. \emph{Read-only} exposure enables immediate harvesting of private data with commodity crawlers, while \emph{editable} exposure or \emph{account access} escalates the harm from confidentiality loss to integrity risk: adversaries can alter user records, submit incorrect information, change account details, or perform actions as the victim. This risk is amplified when these URLs are transmitted over unencrypted channels such as SMS.

\takeaway{We found that 14 services allowed adversaries to modify users' PII, five services that grant account access, and one that gives both account access and edit access. Services should not show any PII without authentication, but they should be especially careful in the case of privileged access.}

\subsection{Potential Downstream Attacks}
\label{subsec:downstream_attacks}
While individual PII attributes can be harmful, their risk is exemplified when multiple identifiers appear together.
Prior work has shown the ability of a group of PII attributes to facilitate downstream attacks, such as spear phishing, social engineering, identity theft, and cross-dataset linkage. For instance, combinations such as name, phone number, email address, postal address, date of birth, financial identifiers, or government IDs can form actionable user profiles. Such contextualized user profiles have great value for targeted marketing and financial fraud. Below, we explore the potential use of the PII leaks in actionable downstream attack pipelines.

\BfPara{\underline{Methodology}}
We look at the UI captures and JSON payloads for each validated URL to extract user PII. Our ground rules are that the URL should expose a user's name (an established and effective identifier) \emph{and} at least one type of contact PII: (1) telephone number, (2) email address, or (3) postal address. If a URL passes these conditions, we extract all visible types of PII. To avoid repetition, we concatenate all visible types of PII for each URL to construct a ``user profile'' and only unique profiles are kept to analyze what types of PII appear together and how some combinations can be more effective than others. Although all pages were publicly reachable, we treat the content as sensitive: plaintext PII values are deleted after analysis and only aggregate counts are stored.

\BfPara{\underline{Results}}
Among the validated set of URLs, 290 (41.37\%) met our ground conditions. After filtering unique ``user profiles'', we were left with 206 (29.39\%) candidates. The broadest PII exposure (large user profile) came from a pre-filled loan application exposing first name, last name, email address, telephone number, date of birth, marital status, postal address, bank account number, state identification card number, and family (refer to appendix Fig.~\ref{fig:wannacash}). Another notable combination was also from a loan application exposing first name, last name, email address, telephone number, postal address, bank account number (including routing number), and social security number in the network logs (refer to appendix Fig.~\ref{fig:epic_loan_systems}).

\BfPara{\underline{Impact}}
The ability of adversaries to aggregate multiple PII attributes (e.g., date of birth, social security number, name) to \textit{profile users} creates avenues for downstream attacks such as spear phishing, social engineering attacks, reconstruction attacks, and identity theft~\cite{das2019sok,syafitri2022social,van2023social,dwork2017exposed, opderbeck2022cybersecurity}. These profiles, even those passing our two ground conditions, can be leveraged to launch downstream threats. For instance, compound profiles with attributes such as {\emph{name}, \emph{email}, \emph{phone}} enable high-confidence spear-phishing and cross-dataset linkage. Similarly, including \emph{DOB} or \emph{account identifiers} intensifies susceptibility to financial risks.

\takeaway{We identified 206 distinct user profiles, demonstrating how SMS-delivered URLs can facilitate social engineering and enable critical downstream attacks, including spear-phishing and financial impacts.}

\subsection{Hidden Data Leaks}
\label{subsec:hidden_data_leaks}
The first page loaded from an SMS-delivered URL may not reveal the full extent of exposed data.
Some services initially display limited information but expose additional PII after simple UI interactions, such as clicking buttons, opening tabs, expanding sections, or navigating to related pages.
In other cases, the UI might have limited or no PII but underlying network requests made by the browser contain additional PII. In many cases, API responses may include additional fields or payloads that are not rendered, yet still remain accessible through browser developer tools (e.g., network logs).
Therefore, we interact with the UI and analyze network logs to identify ``\textit{hidden}'' leaks where the initial page appears to minimize or hide PII, but additional user data is exposed through subsequent navigation or client-side payloads.

\BfPara{\underline{Methodology}}
We examine webpages for navigational elements such as menus, buttons, pagination controls, and expandable accordions. We interact with these elements to uncover additional user PII, avoiding any state-changing actions such as form submissions. Upon the first occurrence of additional PII, we stop and label it as a ``hit'' for UI navigation.
We also investigate whether JSON payloads from HTML and network logs contain additional user PII that is not exposed in the UI, since modern web applications often retrieve data through API responses, JSON payloads, or HTML-bootstrapped state before deciding what to display. To do so, we use the browser profile with DevTools enabled and navigate to the Network tab. For each PII string visible in the UI, we use Network search (body+headers) to find its originating response. Then, we manually skim these responses to find additional user PII not present in the UI.
In addition, we record common patterns and root causes of misconfigurations that lead to overfetching of PII.

\BfPara{\underline{Results}}
We identify hidden data leaks in 84 cases where additional PII is exposed beyond what is visible on the initial page load. Specifically, we identify eight instances where additional PII is exposed through interaction and navigation on the webpage. On the other hand, we observed overfetching in 76 cases due to improper server-side filtering: network responses or server-rendered bootstraps carried more PII than displayed in the UI. Although technologies such as GraphQL support precise data retrieval, we identify three services that use it incorrectly, negating its intended security benefits. Instead of requesting only the fields required for the current view, these services retrieve complete user profiles or order details and then filter the response on the client side.

\noindent \textit{Case Study 1 (EverQuote):}
It allows users to shop for insurance across 160 insurance companies including top providers such as Progressive and Liberty Mutual Insurance. It is consistently ranked around 60k on the Tranco list and the company reports handling ``35 million quote requests from consumers'' \cite{everquoteEverQuote}. One of the URLs in our dataset redirected to an insurance quote page from Everquote. It only showed (“Congratulations, {FirstName}”) with a \textit{Show Quotes} call-to-action (refer to appendix Fig.~\ref{fig:everquote}). However, clicking on the call-to-action button led to a fully editable, pre-filled insurance quote form containing sensitive PII---full name, email address, gender, date of birth, and credit score range. Further, the URL had a short path (token), consisting of five alphabetic characters, which, combined with the large number of users, resulted in a saturated token space. Consequently, we were able to access another user's quote only after three token mutations which also reiterates the cascading effect described in \S\ref{subsec:user_enumeration}.

\noindent \textit{Case Study 2 (DiMOrder):}
It is a POS system for restaurants powering over 1,000 restaurants across Asia. It has received multiple awards including Forbes Asia 100 to Watch and is partnered with major services including Stripe \cite{dimorderDimPOSOnestop}. Therefore, it is a trusted platform used for orders by restaurants. The URL in our dataset showed an order page in the UI with only the user's address for delivery, which was the only information intended to be shown. Still, the JSON response from the API returned the entire order data: seller and user metadata, geolocation, contact details, order statuses, and even the driver's license plate number (refer to appendix Fig.~\ref{fig:dimorder}).

\BfPara{\underline{Impact}}
Although some developers may assume that ``hidden in the UI'' means hidden from attackers, this PII is still accessible from static resources, making it vulnerable to client-side observers. Hidden data leaks expand the risk of SMS-delivered URLs beyond what is visible on the initial page, allowing anyone with the link to uncover additional user information through simple UI interaction or client-side inspection. Even when a landing page appears benign, UI interactions (often in just a few clicks) can reveal additional user PII that was not initially visible. This expands the leak from a single read-only snapshot into an exploratory path accessible to anyone with the link. Similarly, even if the visible page redacts or omits PII, excess payloads in client-side responses and HTML can silently leak sensitive PII. It is also important to note that, in many cases, service providers assume their endpoints expose minimal PII, while hidden UI paths and network logs reveal much more to an attacker.

\takeaway{We found that 84 services expose additional PII beyond the initial page, including 76 through client-side overfetching and 8 through UI interactions or navigation. This is a reminder to protect sensitive UI paths when accessing from an SMS-delivered URL. Further, pages should request only the PII fields needed by the current UI view. During development and testing, server-rendered bootstraps, API/WebSocket responses, and navigable UI paths should be audited for sensitive data.}

\section{Responsible Disclosure}
\label{sec:disclosure}

For each vulnerable service, we manually reviewed its website to identify explicit guidelines for reporting security and privacy weaknesses. During this process, we found disclosure details in privacy policies or custom responsible-disclosure guidelines, while some services declared no disclosure pathway. Our experience highlights the need for organizations to maintain clear reporting instructions, dedicated disclosure channels, and sufficient resources to ensure the timely evaluation and resolution of security reports. The rest of the section describes our disclosure methodology and insights from the experience. Lastly, we describe the impact of our disclosures.

\begin{table}[htb]

  \caption{Responsible Disclosure Status.}  \centering

    \scalebox{0.99}{
    \begin{tabular}{lr}
      \toprule
      \textbf{Disclosure Status}        & \textbf{\#Domains} \\
      \midrule
      Unreachable                      & 27 \\
      No Response                      & 132 \\
      Responded                        & 11 \\
      Fixed                            & 7  \\
      \midrule
      \textbf{Total}                   & 177 \\
      \bottomrule
    \end{tabular}}

  \label{tab:disclosure}
\end{table}

\BfPara{\underline{Disclosure Methodology}} We found that only five services included a dedicated section describing their security policy, and only two of these provided a bug-reporting page. In addition, seven services mentioned privacy-related reporting mechanisms in their privacy policies. For the remaining 165 services, we relied on any publicly available contact channels we could identify. We prioritized email addresses, followed by telephone numbers and web-based contact forms when available. In most cases, the only reachable channels were general customer-service email addresses or phone numbers, rather than contacts for dedicated engineering, security, or privacy teams. For 27 services, we were unable to identify any viable public contact information.
These observations offer insight into the broader security and privacy maturity of the ecosystem. Notably, one service with an explicit security section provided a detailed disclosure form accompanied by a PGP key; however, the key had expired nearly three years ago. In one case, involving a major U.S. sporting goods retailer, the customer-service team did not appear to understand the issue. We spent approximately 15 minutes on the phone explaining the vulnerability, first to a representative and then to a manager. They stated that they would follow up, but that never happened and the vulnerability remains unresolved.

\BfPara{\underline{Disclosure Insights}} We find that responsible disclosure is often a labor-intensive process with a low likelihood of acknowledgment or remediation. Table~\ref{tab:disclosure} describes the disclosure status. Of the 150 reports we submitted, only 18 received responses, and just 7 out of those resulted in implemented fixes. To improve engagement, we adopted an alternative strategy by conducting telephone-based disclosures for services that had neither responded to email nor acted within 30 days.
This approach enabled us to reach teams at four services, three of whom subsequently acknowledged the issues. One notable case involved a provider stating that they were “overwhelmed with the volume of scam bug reports and had assumed our disclosure email was also a scam.” It is important to emphasize that our reporting process adheres to strict standards: each email includes detailed reproduction steps and an accompanying video demonstration or relevant data.

\BfPara{\underline{Disclosure Impact}} We estimate the scale using publicly reported active-user counts, where available, for fixed services that had potentially enumerable tokens. We estimate that the resulting remediations reduced exposure risk for at least 120 million users.

\section{Related Work}

\BfPara{\underline{SMS Abuses}}
Recent studies have focused on understanding security threats originating from SMS messages.
Several works show that SMS metadata and timing patterns can be exploited to infer user locations~\cite{hong2018guti, hussain2019privacy, lakshmanan2021stealthy}.
For example, Bitsikas~\etal~\cite{bitsikas2023freaky} demonstrated that an attacker can deduce the location of an SMS recipient by comparing message-delivery timings against pre-collected measurements from receiver locations, and argued that mitigating this class of attacks requires complex architectural changes on the SMS infrastructure.
Other studies analyzed attacks that use SMS messages as a bridge for spam and phishing  campaigns~\cite{lota2017systematic,murynets2012crime,shafi2017review,reaves2016sending, rahman2023users,timko2023commercial}.
In this regard, Nahapetyan~\etal~\cite{nahapetyan2024sms} collected a longitudinal dataset of phishing SMS messages and categorized them into over 35K phishing campaigns, providing detailed insights into attacker tactics and countermeasures.
Similar to our work, many of these studies utilized public SMS gateways as a way to understand real-world behavior of the SMS messages and their security implications~\cite{nahapetyan2024sms, reaves2016detecting, reaves2018characterizing, moreno2023your, cheng2020characterizing}.
Our work differs in that we focus on benign, service-generated messages containing private, user-specific URLs, and we analyze the design of these URLs and how they can expose user PII.

\BfPara{\underline{URL Security and Privacy}}
Researchers have put effort into discovering potential vulnerabilities, misconfigurations, and leaks in URLs~\cite{matic_identifying_2020,west2014privacy,sahoo2017malicious,krishnamurthy2009leakage}. For instance, Matic~\etal~\cite{matic_identifying_2020} trained a classifier to identify URLs pointing to sensitive user information. Over a database of 155 million sensitive URLs, the authors found questionable practices such as the usage of unencrypted protocols to transfer sensitive data (HTTP). Moreover, Khodayari~\etal~\cite{khodayari_not_2025} analyzed open redirect vulnerabilities and found 623 sites exposed to this issue. The authors revealed that this vulnerability is widespread and affects 9.7\% of the top 10K websites. Similarly, Wang~\etal~\cite{wang_credit_nodat} found overly excessive permissions in cloud-based API endpoints, potentially leading to data attacks. This study was performed over 1.3 million cloud apps, where $\approx27\%$ were exploitable. Complementing these URL-centric studies, Rahaman et al.~\cite{rahaman2019security} evaluated PCI DSS certification for e-commerce websites, revealing substantial gaps between prescribed web-security requirements and their real-world enforcement. Prior work has also shown that low-entropy object identifiers and shortened URLs can permit enumeration and unauthorized discovery of resources \cite{georgiev2016gone, neumann2010security, nikiforakis2011exposing}. Our work does not claim that address-space enumeration itself is novel. Instead, we contextualize a novel attack surface (public SMS gateway) that enables the accurate identification of target services and their token format without relying on blind discovery.

\section{Conclusion}
\label{sec:conclusion}

\noindent Service providers increasingly prioritize frictionless user experiences by adopting technologies that streamline access to their platforms. SMS remains a widely used channel for delivering private, single-click access URLs. However, the inherent insecurity of SMS has led to data leaks that can expose these URLs to unauthorized adversaries. This paper presents a comprehensive examination of the security and privacy implications arising from such unauthorized access to SMS-delivered URLs. Using public SMS gateways as a lens, we analyzed 322K URLs extracted from more than 33 million SMS messages, ultimately identifying expert-validated PII leaks across 701 URLs associated with 177 services. We found 125 services with low-entropy tokens that were potentially enumerable and confirmed enumeration in 13 services under bounded testing (10 manual attempts), demonstrating the potential for exposure to extend beyond the intended recipient. Additionally, we identified 20 services that grant adversaries privileged access, 84 services that expose additional PII beyond the landing page, and 206 unique user profiles that could be exploited for downstream attacks such as spear-phishing. We recommend three concrete defenses:
(i) URLs delivered over SMS should never act as standalone authenticators;
(ii) tokens should provide $\ge64$ bits of entropy from a cryptographically secure RNG;
and (iii) services should never expose more PII than what is needed on the landing page and server-rendered bootstraps, API/WebSocket responses, and navigable UI paths should be audited for sensitive data.

\bibliographystyle{ACM-Reference-Format}
\balance
\bibliography{bib}

\appendix

\section{Open Science}
The artifacts of this paper are publicly available at \url{https://github.com/cs-maestro/cost_of_convenience}. The repository contains scripts and non-sensitive artifacts supporting URL-artifact collection, preprocessing, filtering, PII detection, validation, and analysis. We do not provide SMS-gateway scraping scripts that could be misused to gather messages containing user PII or URLs that lead to user PII. In addition, we do not release raw datasets, SMS messages, URLs, or identifiers that contain, or could be reverse-engineered to expose, private data. We also withhold these materials because we have signed NDAs with some service providers or they have specifically requested anonymity, and because the majority of affected services have not yet remediated the underlying issues.

\section{Ethical Considerations}
\label{sec:ethics}

This study was reviewed and approved by our institution's ethics board (IRB). The primary stakeholders are users whose data appeared in SMS-delivered URLs, other users of vulnerable services, service providers, public SMS gateway operators, and society at large. Although prior work has treated messages on public SMS gateways as public data~\cite{reaves2016sending,reaves2018characterizing}, we treated exposed data as sensitive and minimized interaction with live systems.
We collected messages only from \textbf{public} SMS gateways that exposed them without authentication or technical access controls and used conservative crawlers consistent with prior measurement practices~\cite{hantke2024red,rautenstrauch2024s}. We collected only the minimum data needed to determine whether PII was exposed, and all raw messages, URLs, and screenshots were ultimately deleted at the end of the study. We do not release raw SMS messages, vulnerable URLs, non-redacted screenshots, extracted PII, or SMS-gateway scraping scripts. PII detection, OCR, and LLM analysis were performed locally, so sensitive artifacts were not transmitted to third-party analysis services. Enumeration and authentication testing was also strictly bounded. Confirmed vulnerabilities were disclosed through providers' published contact or security channels, with at least 45 days provided for remediation.

\section{Public SMS Gateways}
\label{app:gateways}
\noindent azrotv.com, bestsms.xyz, clearcode.cn, freeonlinephone.org, freephonenum.com, getsms.cc, lothelper.com, zh.mytrashmobile.com, receive-a-sms.com, receivefreesms.net, receive-sms.cc, receivesms.cc, receivesms.co, receive-sms-free.net, receivesmsonline.net, receivesmsonline.in, receive-smss.com, shejiinn.com, smstome.com, storytrain.info, temp-number.com, temporary-phone-number.com, us-phone-number.com, xinghai.party, zusms.com

\section{PII Taxonomy}
\label{app:pii_taxonomy}
\noindent mothers maiden name, password, security question, user name, products purchased, psychological trends, search history, services purchased, email address, email content, postal address, telephone number, text messages, citizenship, color, date of birth, education, employment history, family, first name, gender, gender identity, hair color, height, language, marital status, military or veteran status, place of birth, race, religion, sex life, union membership, immigration status, last name, bank account number, credit card number, debit card number, financial account number, insurance policy number, driver authorization card number, health insurance identification number, individual taxpayer identification number, medical identification number, military identification card number, nondriver state identification card number, passport number, social security number, state identification card number, drivers license number, aids, breastfeeding, cancer, childbirth, chiropractic, diagnosis, disability, health records, medicine, mental condition, pregnancy, genetic data, hiv, request for family care leave, request for leave for an employees own serious health condition, location, records of personal property

\section{Case Study Figures}
\label{app:case_studies_figs}

All vulnerabilities illustrated in this appendix were responsibly disclosed and patched before publication. We include only remediated cases and redact all user-identifying information.

\begin{figure}[H]
\centering
\includegraphics[width=0.9\linewidth]{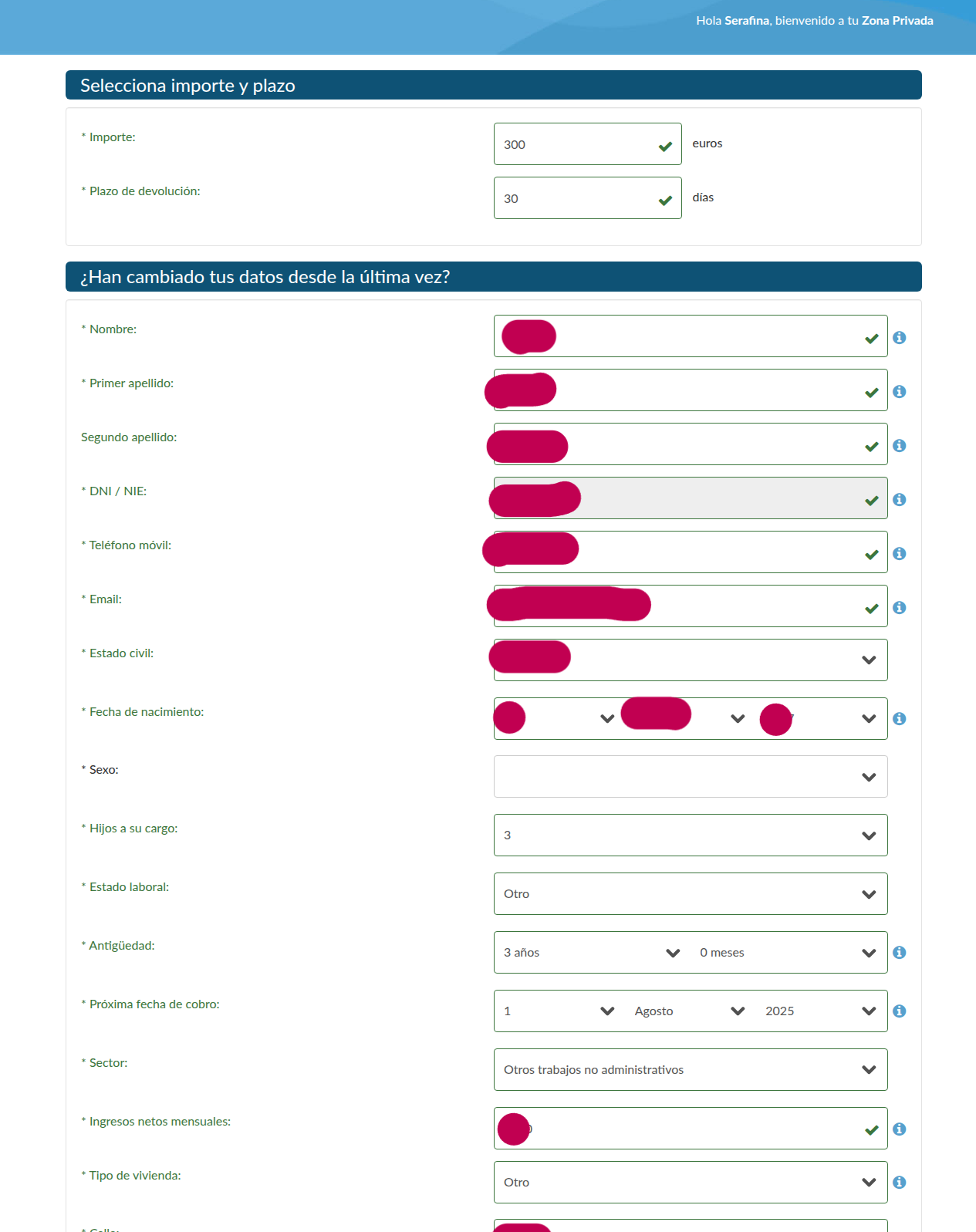}\hfill
\caption{Anonymized case study}
\Description{Redacted screenshot of an anonymized pre-filled application illustrating exposure of multiple user PII fields.}
\label{fig:wannacash}
\end{figure}

\begin{figure}[H]
\centering
\includegraphics[width=0.9\linewidth]{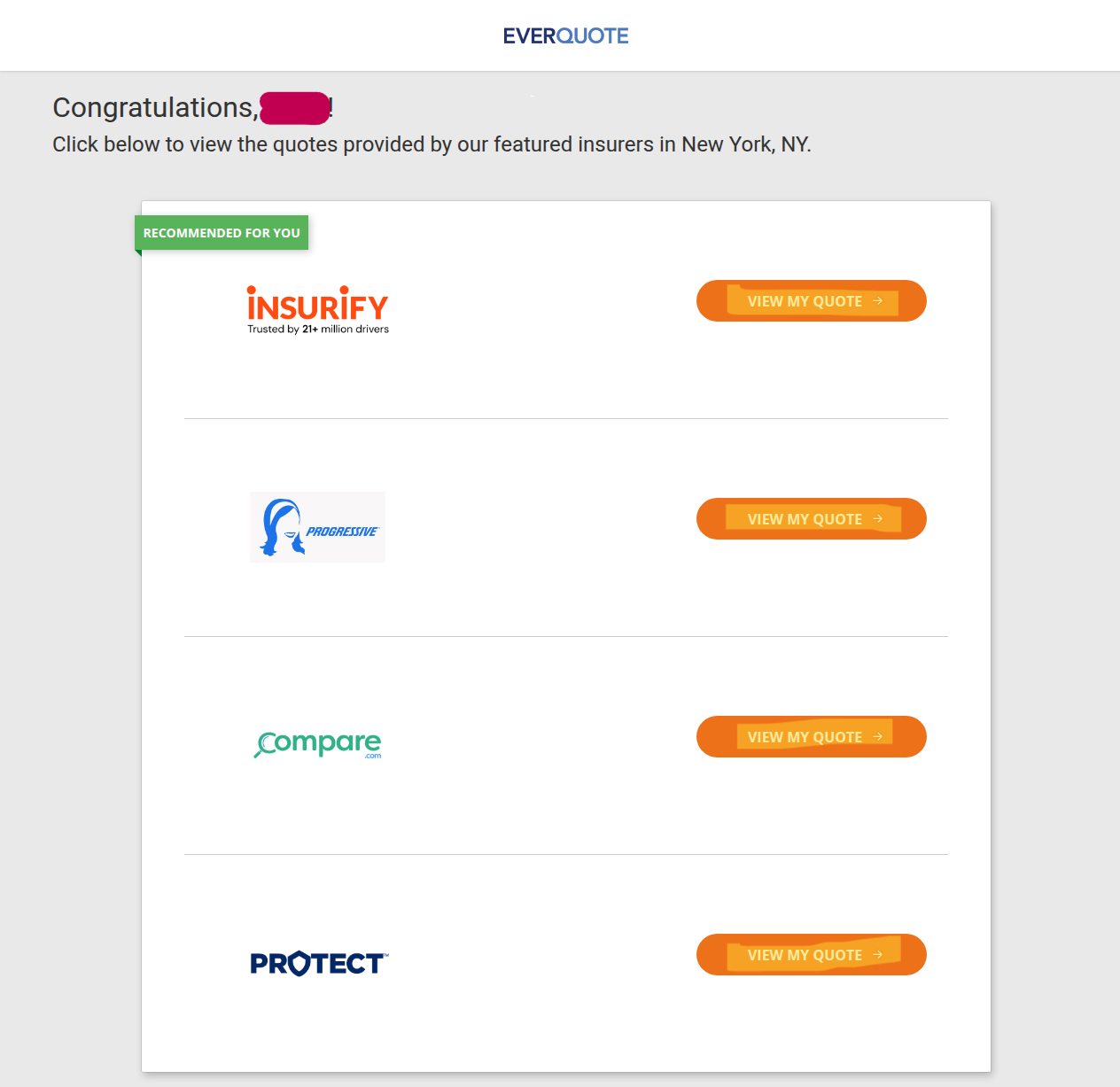}\hfill
\caption{everquote.com}
\Description{Redacted screenshot of an EverQuote insurance-quote flow illustrating user-specific information exposed after following an SMS-delivered link.}
\label{fig:everquote}
\end{figure}

\begin{figure}[H]
\centering
\includegraphics[width=0.9\linewidth]{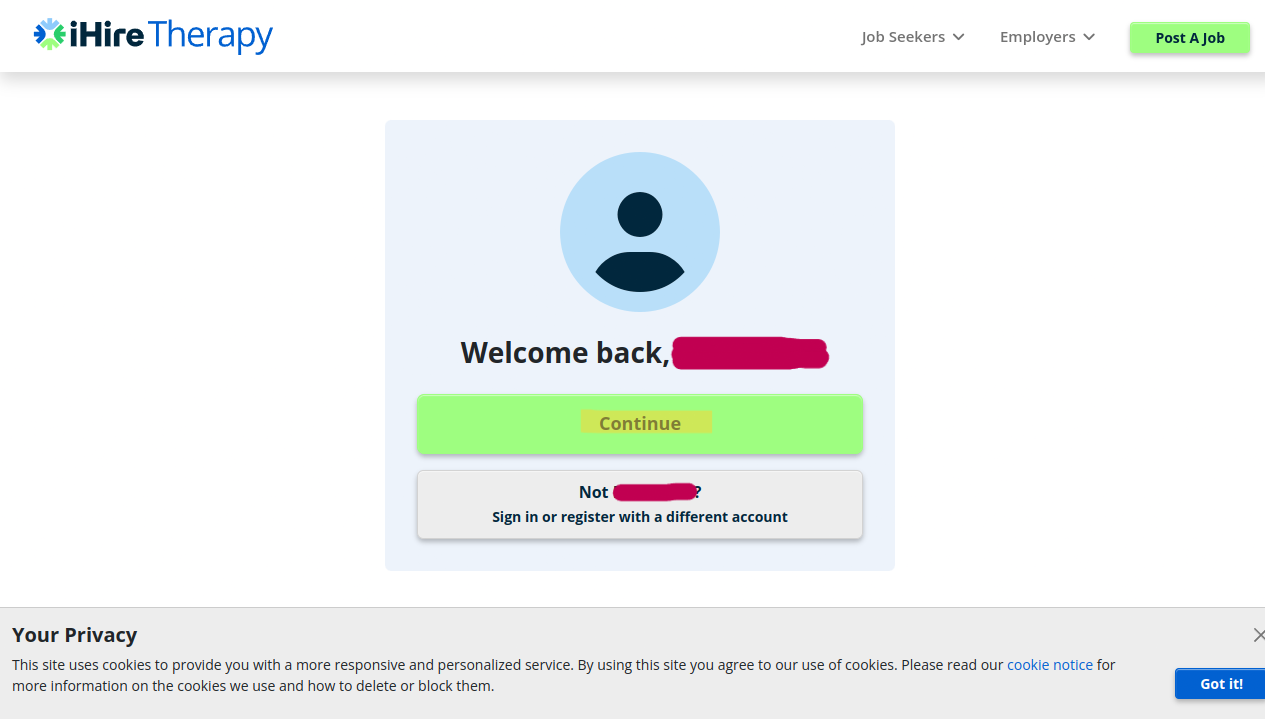}\hfill
\caption{ihire.com}
\Description{Redacted screenshot of the iHire flow illustrating account access reached from an SMS-delivered link.}
\label{fig:ihire}
\end{figure}

\begin{figure}[H]
\centering
\includegraphics[width=0.9\linewidth]{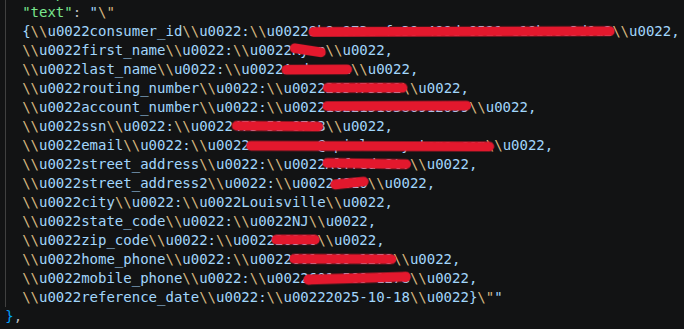}\hfill
\caption{Anonymized case study}
\Description{Redacted screenshot and browser developer-tools view of an anonymized lending application illustrating sensitive user data exposed in client-side responses.}
\label{fig:epic_loan_systems}
\end{figure}

\begin{figure}[H]
\centering
\includegraphics[width=0.9\linewidth]{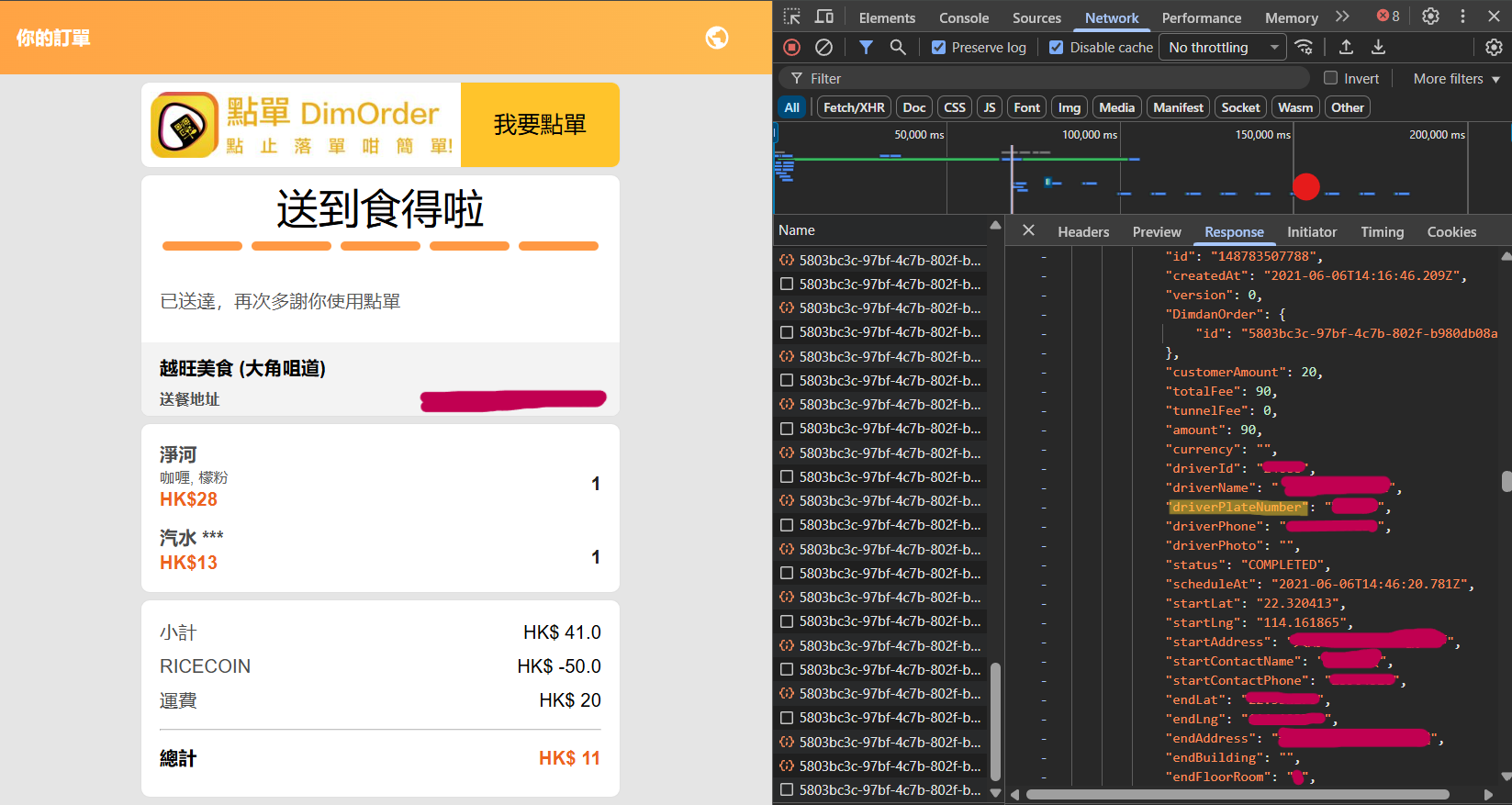}\hfill
\caption{dimorder.com}
\Description{Redacted DimOrder order page and browser developer-tools view illustrating additional user and order metadata returned in a JSON response beyond what is shown in the interface.}
\label{fig:dimorder}
\end{figure}

\begin{figure}[H]
\centering
\includegraphics[width=0.9\linewidth]{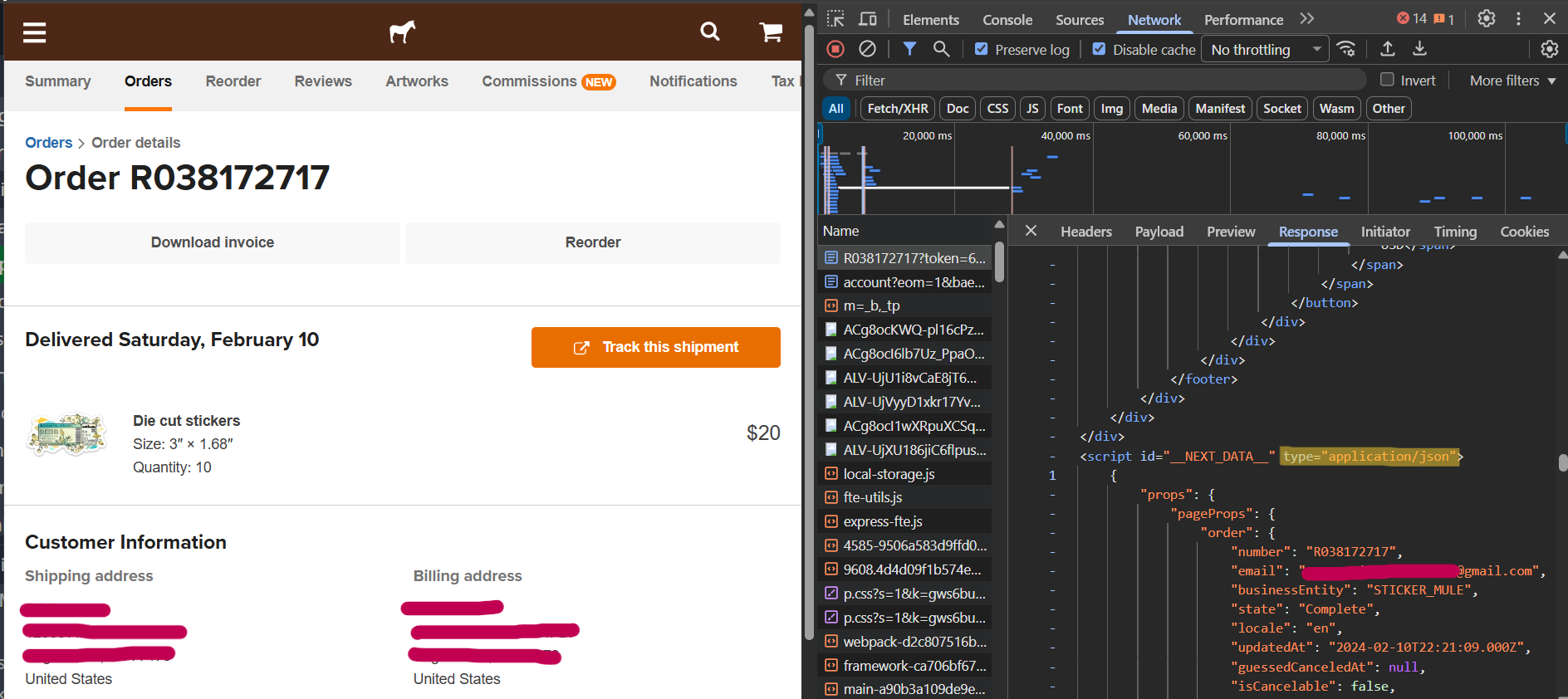}\hfill
\caption{stickermule.com}
\Description{Redacted Sticker Mule order view and associated client-side data illustrating user information linked to an order.}
\label{fig:stickermule}
\end{figure}

\end{document}